\documentclass{caosp306}
\usepackage{graphicx}

\usepackage{natbib}
\articleNo{0}
\pubyear{0}
\volume{0}
\volnumber{0}
\firstpage{0}
\received{\ldots}
\accepted{\ldots}

\def\BibTeX{{\rm B\kern-.05em{\sc i\kern-.025em b}\kern-.08em
             T\kern-.1667em\lower.7ex\hbox{E}\kern-.125emX}}
\def\eprint{e-print:}
\begin{document}

%%%%%%%%%%%%%%%%%%%%%%%%%%%%%%%%%%%%%%%%%%%%%%%%%%%%%%%%%%%%%%%%%%%%%%%%%%%%%
%              R U N N I N G   P A G E   H E A D I N G S                     
% Odd page headings (except for the title page) are produced automatically
% and contain the title. If, and only if, the title of your article is too
% long the running head is omitted in the printout; you can make your own
% running title by using the \htitle command, putting the shortened title
% between the curly brackets. \htitle should also be used when the
% subtitle is present: \htitle offers you a way how to include it into the
% headings. If you wish to see how it works simply remove the % sign from
% the beginning of that line.
%
% Unlike the \htitle command, the \hauthor command is compulsory. It is
% used to produce even page headings and contains the names of the authors
% of an article.  All authors must be listed here, if possible. When
% authors' list is too long, you can abbreviate it by using "{\it et
% al.}". Authors' names are given in the form: initial(s) of the author's
% first name and surname. Authors are separated by a "," (comma) sign and
% the last one by "and".
%%%%%%%%%%%%%%%%%%%%%%%%%%%%%%%%%%%%%%%%%%%%%%%%%%%%%%%%%%%%%%%%%%%%%%%%%%%%%
%\htitle{A note to comet ejection process ...}
\hauthor{P. Marziani et al.}
%\hauthor{L.\,Neslu\v{s}an {\it et al.}}

%%%%%%%%%%%%%%%%%%%%%%%%%%%%%%%%%%%%%%%%%%%%%%%%%%%%%%%%%%%%%%%%%%%%%%%%%%%%%
%                       T I T L E                                            
% Capital letters in the title are only used at the beginning of the
% names. Don`t end the title by a "." (dot)
%%%%%%%%%%%%%%%%%%%%%%%%%%%%%%%%%%%%%%%%%%%%%%%%%%%%%%%%%%%%%%%%%%%%%%%%%%%%%
\title{Quasar emission lines as virial luminosity estimators}

%%%%%%%%%%%%%%%%%%%%%%%%%%%%%%%%%%%%%%%%%%%%%%%%%%%%%%%%%%%%%%%%%%%%%%%%%%%%%
%                       S U B T I T L E                                      
% You can use the subtitle, with the command \subtitle similar to the
% \title command.
%%%%%%%%%%%%%%%%%%%%%%%%%%%%%%%%%%%%%%%%%%%%%%%%%%%%%%%%%%%%%%%%%%%%%%%%%%%%%

%%%%%%%%%%%%%%%%%%%%%%%%%%%%%%%%%%%%%%%%%%%%%%%%%%%%%%%%%%%%%%%%%%%%%%%%%%%%%
%                   A U T H O R  N A M E S                                   
% Authors' names are separated by the \and command and their institutes
% are assigned by the \inst{n} command.
%
% When the name contains "Slovak" letters L,d,t,l with a caron, use an
% a new \softl, etc. command (examples given in the last table of
% this document) to produce typographically correct accented characters.
%%%%%%%%%%%%%%%%%%%%%%%%%%%%%%%%%%%%%%%%%%%%%%%%%%%%%%%%%%%%%%%%%%%%%%%%%%%%%
\author{
        P. Marziani \inst{1}  
      \and 
        E. Bon    \inst{2}   
      \and 
        N. Bon \inst{2}
        \and 
      M. L. Martinez-Aldama\inst{3}
      \and    
        G. M. Stirpe\inst{4}
        \and 
        M. D'Onofrio\inst{5}
        \and A. del Olmo\inst{6}
        \and C. A. Negrete\inst{7}
        \and D. Dultzin\inst{7}
       }

%%%%%%%%%%%%%%%%%%%%%%%%%%%%%%%%%%%%%%%%%%%%%%%%%%%%%%%%%%%%%%%%%%%%%%%%%%%%%
%           I N S T I T U T E S'  A D D R E S S E S                          
% The affiliation of authors is generated by the \institute command, the
% \and command being again used to separate individual addresses.
% The following commands may be used for the following three institutes:   
%               \lomnica        for      AsU SAV, Tatranska Lomnica          
%               \blava          for      AsU SAV, Bratislava                 
%               \ondrejov       for      AsU CAV, Ondrejov                   
%
% The given postal address must be complete in order to facilitate our
% editorial work. Moreover, you can add your e-mail address, using the
% \email command.
%%%%%%%%%%%%%%%%%%%%%%%%%%%%%%%%%%%%%%%%%%%%%%%%%%%%%%%%%%%%%%%%%%%%%%%%%%%%%
\institute{
          National Institute for Astrophysics (INAF), Padua Astronomical Observatory, Padua,  \email{paola.marziani@inaf.it}\\
          Italy
         \and 
          Astronomical Observatory, Belgrade, Serbia
         \and
         Center for Theoretical Physics, Polish Academic of Sciences, Warsaw, Poland
        \and 
        INAF - Osservatorio Astrofisica e Science dello Spazio, Bologna, Italy
        \and
        Dipartimento di Fisica ed Astronomia ``Galileo Galilei,'' Universit\`a di Padova, Padova, Italia
         \and
         Instituto de Astrofisica de Andalucia (CSIC), Granada, Spain
         \and
         Instituto de Astronomia, UNAM, Mexico, D.F., Mexico
          }

%%%%%%%%%%%%%%%%%%%%%%%%%%%%%%%%%%%%%%%%%%%%%%%%%%%%%%%%%%%%%%%%%%%%%%%%%%%%%
%                        D A T E / R E C E I V E D                          
% Date inserted here will be the date when your paper was received The
% format is: month (not abbreviated), day, year.
%%%%%%%%%%%%%%%%%%%%%%%%%%%%%%%%%%%%%%%%%%%%%%%%%%%%%%%%%%%%%%%%%%%%%%%%%%%%%
\date{\ldots}
%\date{March 10, 2003}

%%%%%%%%%%%%%%%%%%%%%%%%%%%%%%%%%%%%%%%%%%%%%%%%%%%%%%%%%%%%%%%%%%%%%%%%%%%%%
%                        M A K E T I T L E
% The beginning part (title, author(s), etc.) of your article must be
% closed by the \maketitle command.
%%%%%%%%%%%%%%%%%%%%%%%%%%%%%%%%%%%%%%%%%%%%%%%%%%%%%%%%%%%%%%%%%%%%%%%%%%%%%
\maketitle

%%%%%%%%%%%%%%%%%%%%%%%%%%%%%%%%%%%%%%%%%%%%%%%%%%%%%%%%%%%%%%%%%%%%%%%%%%%%%
%                        A B S T R A C T,  K E Y W O R D S                   
% Here it is shown how to write an abstract.  Keywords should be placed
% within the "abstract" environment using the command \keywords and they
% should be selected from the thesaurus from Astron.  Astrophys.
% Abstracts. They must be separated from each other by -- (two dashes).
%%%%%%%%%%%%%%%%%%%%%%%%%%%%%%%%%%%%%%%%%%%%%%%%%%%%%%%%%%%%%%%%%%%%%%%%%%%%%
\begin{abstract}
Quasars accreting matter at very high rates (known as extreme Population A [xA]) may provide a new class of distance indicators covering cosmic epochs from  present day up to less than 1 Gyr from the Big Bang. We report on the developments of a method that is based on  ``virial luminosity'' estimates from measurements of emission line widths of xA quasars. The approach is conceptually  equivalent to the virial estimates based on early and late type galaxies.  The main issues related to the cosmological application of luminosity estimates  from xA quasar line widths are the identification of proper emission lines whose broadening is predominantly virial over a wide range of luminosity, and the assessment of the effect of the emitting region orientation with respect to the line of sight. We report on recent developments concerning the use of the AlIII$\lambda$ 1860 intermediate ionisation line and of the Hydrogen Balmer line H$\beta$ as ``virial broadening estimators.''  

\end{abstract}

%%%%%%%%%%%%%%%%%%%%%%%%%%%%%%%%%%%%%%%%%%%%%%%%%%%%%%%%%%%%%%%%%%%%%%%%%%%%%
%                       S E C T I O N I N G                                  
% Any section starts with the command \section as shown below, with the
% title in Initial Capitals and lowercase only. Do not number the sections
% - let LaTeX do that for you - and do not end them by a "." (dot).
%
% The (sub)section titles are typeset in boldface; so, if working in the
% mathematics mode in (sub)section titles, you must use \boldmath and 
% enclose it into curly brackets, e.g. "{\bolmath $R^{2}$}".
%%%%%%%%%%%%%%%%%%%%%%%%%%%%%%%%%%%%%%%%%%%%%%%%%%%%%%%%%%%%%%%%%%%%%%%%%%%%%
\section{Introduction: a main sequence for quasars}
%%%%%%%%%%%%%%%%%%%%%%%%%%%%%%%%%%%%%%%%%%%%%%%%%%%%%%%%%%%%%%%%%%%%%%%%%%%%%
%                       L A B E L                                            
% The label command is very convenient for you when referring to sections,
% subsections,..., tables, figures as well as to equations (see commands
% \ref and \pageref). In the case of figure and/or table environments the
% \label command should always be put after the \caption command to
% preserve proper numbering. When using the \label command the file must
% be compiled twice to get proper cross-references.
%%%%%%%%%%%%%%%%%%%%%%%%%%%%%%%%%%%%%%%%%%%%%%%%%%%%%%%%%%%%%%%%%%%%%%%%%%%%%
\label{intr}
 %%%%%%%%%%%%%%%%%%%%%%%%%%%%%%%%%%%%%%%%%%%%%%%%%%%%%%%%%%%%%%%%%%%%%%%%%%%%%
%                       P A R A G R A P H                                    
% To generate a paragraph simply leave a blank line after the last
% sentence of the preceding paragraph as shown below.
%%%%%%%%%%%%%%%%%%%%%%%%%%%%%%%%%%%%%%%%%%%%%%%%%%%%%%%%%%%%%%%%%%%%%%%%%%%%%

The Main Sequence (MS)   is a powerful tool to organize type-1 quasar diversity \citep[see e.g.,][for a recent review]{marzianietal18}. The MS concept originated from the first eigenvector (E1) of a Principal Component Analysis carried out on a sample of $\approx$80 Palomar-Green quasars \citep{borosongreen92}.  The E1 MS was first associated {\bf with anti-correlations between  strength of FeII$\lambda$4570 and  width of H$\beta$\ as well as strength of FeII$\lambda$4570 and [OIII] prominence}. Over the years, several  parameters related to the accretion process and the accompanying outflows were found to be also associated with the fundamental relation between prominence of singly-ionized iron emission and broad Balmer line width \citep{sulenticetal00a,sulenticetal07}. For an exhaustive list see \citet{sulenticetal11,fraix-burnetetal17}. Since 1992,  the E1 MS has been found in increasingly larger samples, and the MS  potential was fully recognized following an SDSS-based analysis of a large sample of several tens of thousands of quasars \citep{shenho14}.
 
\begin{figure}
\centerline{\includegraphics[width=0.75\textwidth,clip=]{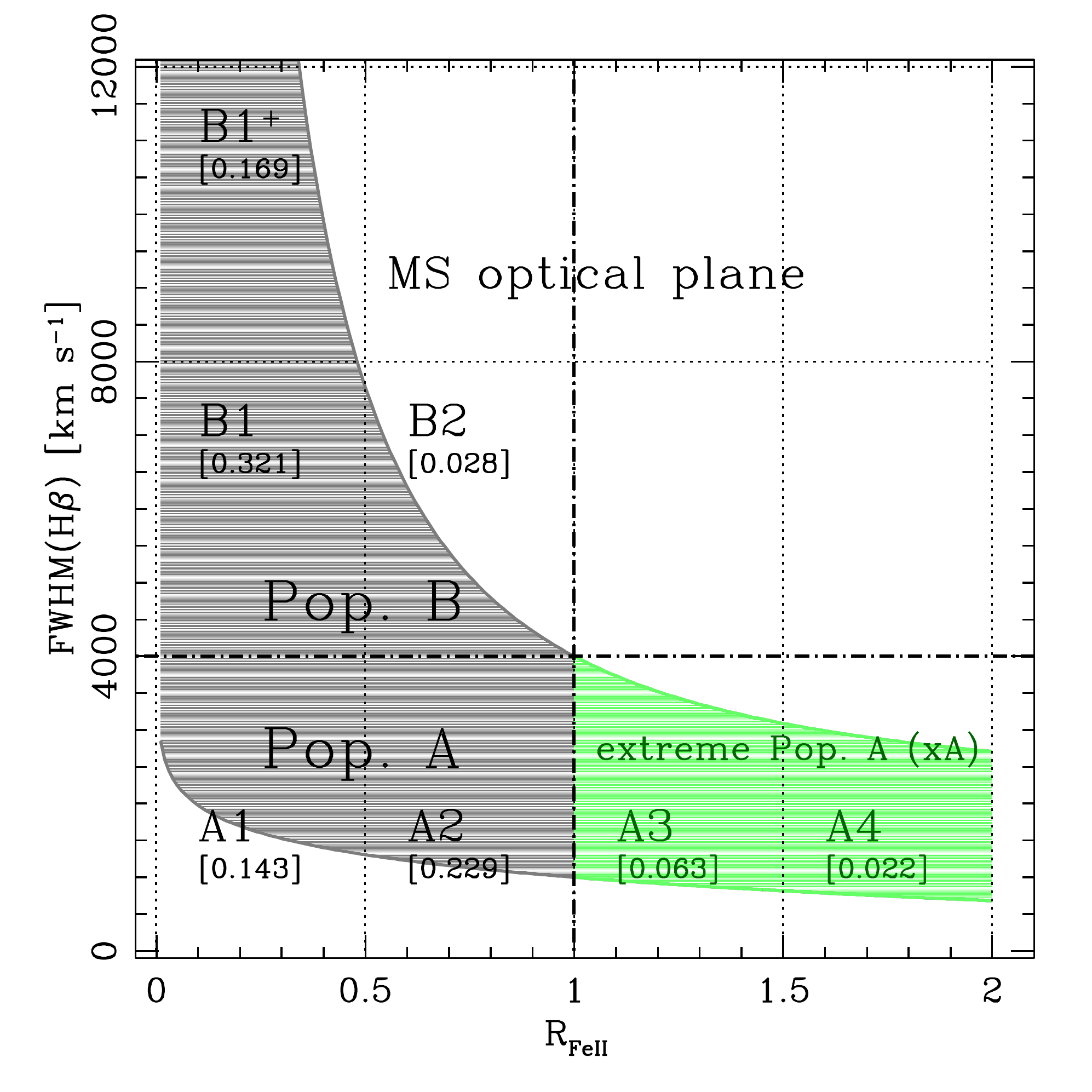}}
\caption{A sketch outlining the occupation of optical plane of the quasar MS. The plane has been subdivided in spectral types, and the region of extreme Population A has been shaded green. The numbers in square brackets are the relative prevalence in the ST along the sequence, from the sample of \citet{marzianietal13a}. }
\label{fig:ms}
\end{figure}

The optical plane of the MS is defined by the FWHM of the $H\beta$\ broad component vs the parameter $R_\mathrm{FeII}$\ that is defined as $R_\mathrm{FeII} =$ I(FeII$\lambda$4570)/I(H$\beta$), the ratio between the integrated flux of the FeII$\lambda$4570 blend of multiplets, and that of the H$\beta$ \  broad component \citep{sulenticetal00a}. The data are distributed  as an elbow-shaped figure (Fig. \ref{fig:ms}) if a restriction to low-$z$ and relatively low luminosity is applied; if higher luminosity sources are included the MS takes a wedge-like form (as in case of \citealt{shenho14}).  

It is not surprising that a measure of  FeII emission can lead a fundamental correlation, as singly-ionized emission is extended from the UV to the NIR, and  can dominate the thermal balance of the {\bf broad line region }(BLR)  \citep{marinelloetal16}. FeII emission is self-similar but its intensity with respect to Hydrogen Balmer line H$\beta$  changes. The width of H$\beta$\ is mainly associated with the  broadening due to the emitting gas dynamics via Doppler effect.  A correlation between the two parameters points towards a coupling between physical and dynamical conditions within the line emitting region, as further discussed in \S\ \ref{vw}.  

Quasar spectra show a wide range of line profiles, line shifts, line intensities and differences in dynamical conditions and ionization levels of the BLR, systematically changing along the sequence. It is expedient to identify spectral types along the  MS  (Fig. \ref{fig:ms}) and to distinguish  between Population A (FWHM(H$\beta$) $ \le$ 4000 km s$^{-1}$) and Population B {\bf (FWHM(H$\beta$) $>$ 4000 km s$^{-1}$)}. Pop. A sources include prototypical Narrow Line Seyfert 1 IZw1 as well as sources with relatively little FeII, such as Mark 335. The appearance of the spectrum suggests a relatively low degree of ionization, with significant FeII, weak [OIII]$\lambda\lambda$\-4959,5007 and weak high-ionization lines in general. Pop. B objects show not only broader lines but also higher ionization, weak FeII and strong [OIII]. A prototypical source is NGC 5548 \citep{sulenticetal00a}. Regarding internal line shifts in the spectra of individual quasars, it is helpful  to distinguish between low- and high-ionization lines (HILs and LILs). Internal line  shifts between HILs and LILs are mainly associated with HIL blueshifted emission.   

The MS   is probably due to a combination of effects dependent on Eddington ratio, viewing angle, and metal content (\citealt{pandaetal19} and references therein).  The connection between observational and accretion parameters is not well-mapped as yet, but the quasar MS could be aptly considered as the analogous of the MS in the stellar H-R diagram  \citep{sulenticetal01,sulenticetal08}.
 
Even if our understanding is not complete, it is possible to exploit some well-defined properties in particular sectors ``spectral types" along the MS. A second key element that makes it possible to consider quasars even as ``Eddington standard candles" (i.e., sources with a small scatter around a fixed value of the Eddington ratio in place of a luminosity measure) has been the ability to recognize that the part of the BLR emitting the LIL is eminently virialized. In the following we will  summarize in which way quasar emission lines can be considered as virial broadening estimators (VBE, \S \ref{vw}) and identify a particular  class of quasars (\S \ref{xa}) for which the assumption of an almost constant Eddington ratio is likely to be verified. We  stress the analogy with stellar systems (\S \ref{ss}) and consider redshift-independent estimates of luminosity made possible by the scaling between virial broadening and luminosity itself (\S \ref{cs}).

\section{A virialized and a wind system} 
\label{vw}	

\begin{figure}
\centerline{\includegraphics[width=0.75\textwidth,clip=]{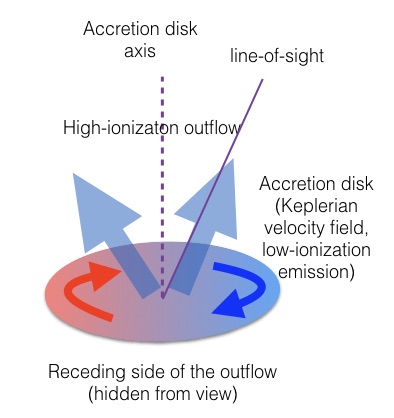}}
\caption{A sketch illustrating the principle of a virialized sub-region co-existing with an outflowing component. The sketch is highly simplified and  accounts for the systematic blueshifts of HILs, large when the HI Balmer lines are narrower, but does not takes into account the systematic changes in accretion modes expected along the sequence.}
\label{fig:vw}
\end{figure}

The origin of the broadening of quasar emission lines has been a contentious issue for decades after quasar discovery and it is as yet not fully understood.  Reverberation-mapping campaign have provided direct evidence of a large mass concentration in a very small volume of space, with broadening for lines emitted by different ionic species becoming larger and their distance decreasing with increasing ionisation potential \citep{petersonwandel99}. At the same time, internal line shifts (measured soon after the quasar discovery, \citealt{burbidgeburbidge67}) complicated the interpretation of  the spectra of individual quasars. A turning point was the ability offered by the FOS on board HST to observe the HIL CIV 1549 and to compare it to a strong LIL such as H$\beta$ \citep{corbinboroson96,marzianietal96}. It was found that the CIV line showed prominent blueshift while  the   H$\beta$  remained almost unshifted with respect to the rest frame of the quasars.  If we assume that  a line whose profile appears symmetric and unshifted with respect to rest frame can be considered as a marker of a virialized emitting region,  this result provided support to the idea that the BLR were composed of two sub-regions, one emitting  predominantly HIL in an outflow scenario and one associated with a flattened distribution of gas coplanar with the accretion disk (possibly the accretion disk itself, \citealt{collinsouffrinetal88,elvis00}; see Fig. \ref{fig:vw}). Later, it was found that a virialized system emitting mainly LILs coexists with outflowing gas in Pop. A sources, even at the highest luminosity  \citep{sulenticetal17,vietrietal17,coatmanetal17}.

%\citep{elvis00} % ; Dultzin-Hacyan et al. 1997; Shang et al. 2003, Yip et al.2004, Sulentic et al. 2000, 2002, 2007; Kruzcek et al 2011; Tang et al. 2012; Kuraszkiewicz et al. 2008; Mao et al. 2009; Grupe 2004, Wang et al. 2006; Panda et al. 2018, 2019. SDSS data : Shen & Ho 2014, Sun & Shen 2015, Brotherton et al. 2015; Du et al. 2016.

\begin{figure}
\centerline{\includegraphics[width=1.0\textwidth,clip=]{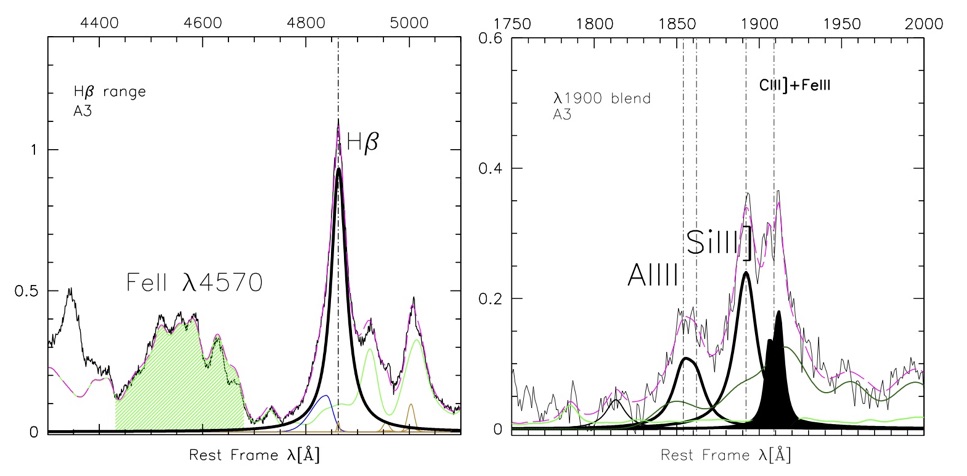}}
\caption{Examples of the H$\beta$\ (left, {\bf \citealt{marzianietal13}})  and 1900 \AA\ (right, \citealt{bachevetal04})  spectral region (after continuum subtraction) in low  redshift extreme Eddington candidates, selected from the optical criterion $R_\mathrm{FeII} > 1 $.  The   ratio  $R_\mathrm{FeII}$ is obtained by measuring the flux of the {Fe}{II} blend at 4570 \AA\ (pale green shaded area), and the flux of H$\beta$.  The { thick} black lines  trace   broad lines whose FWHM can be used as virial broadening estimator,   {Al}{III}$\lambda$1860, {Si}{III}]$\lambda$1892, and H$\beta$\  but not {C}{III}]$\lambda$1909 (shaded dark). H$\beta$\ and the UV lines  have been found to  provide consistent FWHM measures in agreement with H$\beta$\  at low-$L$.  Green lines show  Fe{\sc ii} templates  (pale), and the Fe{\sc iii} template in the UV  (dark). The magenta dashed lines show models of the total line emission.}
\label{fig:spectra}
\end{figure}

\section{ Extreme Population A}
	\label{xa}

Selection and physical conditions of  Extreme Population A is made possible by the MS that allows for the definition of spectral types, in addition to the two Populations defined earlier. Spectral types within Pop. A are due to  a gradient of $R_\mathrm{FeII}$  \citep{borosongreen92,sulenticetal00a,shenho14,duetal16a,pandaetal19}. 

Extreme Pop. A quasars (xA) are selected applying simple  criteria from diagnostic emission line intensity ratios:
\begin{enumerate} 
\item $R_\mathrm{FeII}$  FeII$\lambda$4570 blend/H$\beta$ $\ga$ 1.0;
\item UV AlIII $\lambda$1860/\-SiIII]$\lambda$1892$\ga$\-0.5 and SiIII]$\lambda$1892/\-CIII]$\lambda$1909$\ga$1
\end{enumerate} 

 UV and optical selection criteria are equivalent, and lead to the identification as xA of
 $\sim$ 10\% of all quasars in low-$z$, optically selected samples. xA spectra show distinctively strong FeII emission and Lorentzian Balmer line profiles, and  FWHM(H$\beta$) $\approx$ FWHM(AlIII 1860) whenever it has been possible to cover both lines for the same object (\citealt{negreteetal13,marzianisulentic14}). This means that the AlIII 1860 FWHM is a  virial broadening estimator equivalent to H$\beta$\ (del Olmo et al. 2019, in preparation).  The measurements of  the 1900 blend  and  specifically of the AlIII line  allows for the consideration of xA quasars at high $z$ ($\ga1.2$). 
 
The UV spectrum of xA quasars at $z \sim 2$ shows symmetric low-ionization and blueshifed high- ionization lines even at the highest luminosity (\citealt{martinez-aldamaetal18}, and reference therein). The CIV and the 1900 blend lines have low equivalent width: about one-half of xAs are weak lined quasars following the definition of \citet{diamond-stanicetal09}.   

Observed diagnostic ratios (CIV/AlIII, CIV/HeII, AlIII/SiIII]) imply extreme values for density (high, $n \ga  10^{12-13}$ cm$^{-3}$), ionization (low, $U \sim 10^{-3}$) and extreme values of metallicity ($Z\ga 20 Z_{\odot}$, \citealt{martinez-aldamaetal18},  Sniegowska et al. 2019 in preparation). These values are inferred by considering curves of constant intensity ratios in the  plane ionization parameter versus density from arrays of CLOUDY simulations \citep{negreteetal12}. A crossing point defined by the isopleth of constant intensity ratios identifies the physical conditions that predict all considered intensity ratios. The very high density inferred from the weakness of CIII]$\lambda$1909, and the maximum radiative output per unit mass suggests that the LIL Emitting regions could be a  dense compact remnant of a more conventional LIL-BLR.

\section{Extreme Population A quasars: virial broadening}
\label{vl}

%xA: Marziani & Sulentic 2014 (MS14); Negrete et al. 2018; Martinez-Aldama et al. 2018; 
  There are several lines of evidence suggesting that the $R_\mathrm{FeII}$ is correlated with Eddington ratio. Among them, we mention the fundamental plane of the accretion black holes correlations \citep{duetal16a} which is a restatement of the correlation of Eddington ratio $R_\mathrm{FeII}$, adding a second correlate associated with the LIL profile.  Extreme Population A ($R_\mathrm{FeII}$ $\ga$ 1) show  extreme $L/L_\mathrm{Edd}$ along the MS\ with small dispersion \citep{marzianisulentic14} (MS14). \  Accretion disk theory predicts low radiative efficiency at high accretion rate; $L/L_\mathrm{Edd}$ saturates toward a limiting values; \citep{mineshige00,abramowiczetal88,sadowski11}. This seems to be what is occurring to xA sources. 

A virial luminosity estimate for large samples of xA quasars  is possible if: 
\begin{enumerate}
\item xA quasars radiate close to Eddington limit $L/L_\mathrm{Edd} \propto L/M_\mathrm{BH} \sim 1 $. The exact value of the the Eddington ratio is not relevant, provided that   $L/L_\mathrm{Edd}$  scatters little around a well-defined value. At this point we are able to recognize only xA quasars at one extreme of the MS, but the same approach could be applied to other sources along the MS, were their Eddignton ratio known with high precision. 
\item   Virial motions are the broadening source of the low-ionization BLR.
\item xA quasars have similar BLR physical parameters, an assumption that is justified by the spectral similarity over a large range in luminosity.  This implies that BLR radius rigorously scales with the square root of the luminosity.
\end{enumerate}

If we know a virial broadening estimator $\delta v$ (in practice, the FWHM of a low-
ionization line), we can derive a $z$-independent, ``virial'' luminosity, $L_\mathrm{vir} \propto  (n_\mathrm{H} U)^{-1} (\delta v)^{4}$. The virial luminosity can be written as $L = {\mathcal L_{0}} $ FWHM$^{4}$, where the constant depends on the fraction of ionizing luminosity, the  average frequency of ionizing photons, and the photon flux. These are all intrinsic properties of quasars.

\subsection{$L \propto (\delta v)^{a}$: not only for quasars}
\label{ss}

\begin{figure}
\centerline{\includegraphics[width=0.75\textwidth,clip=]{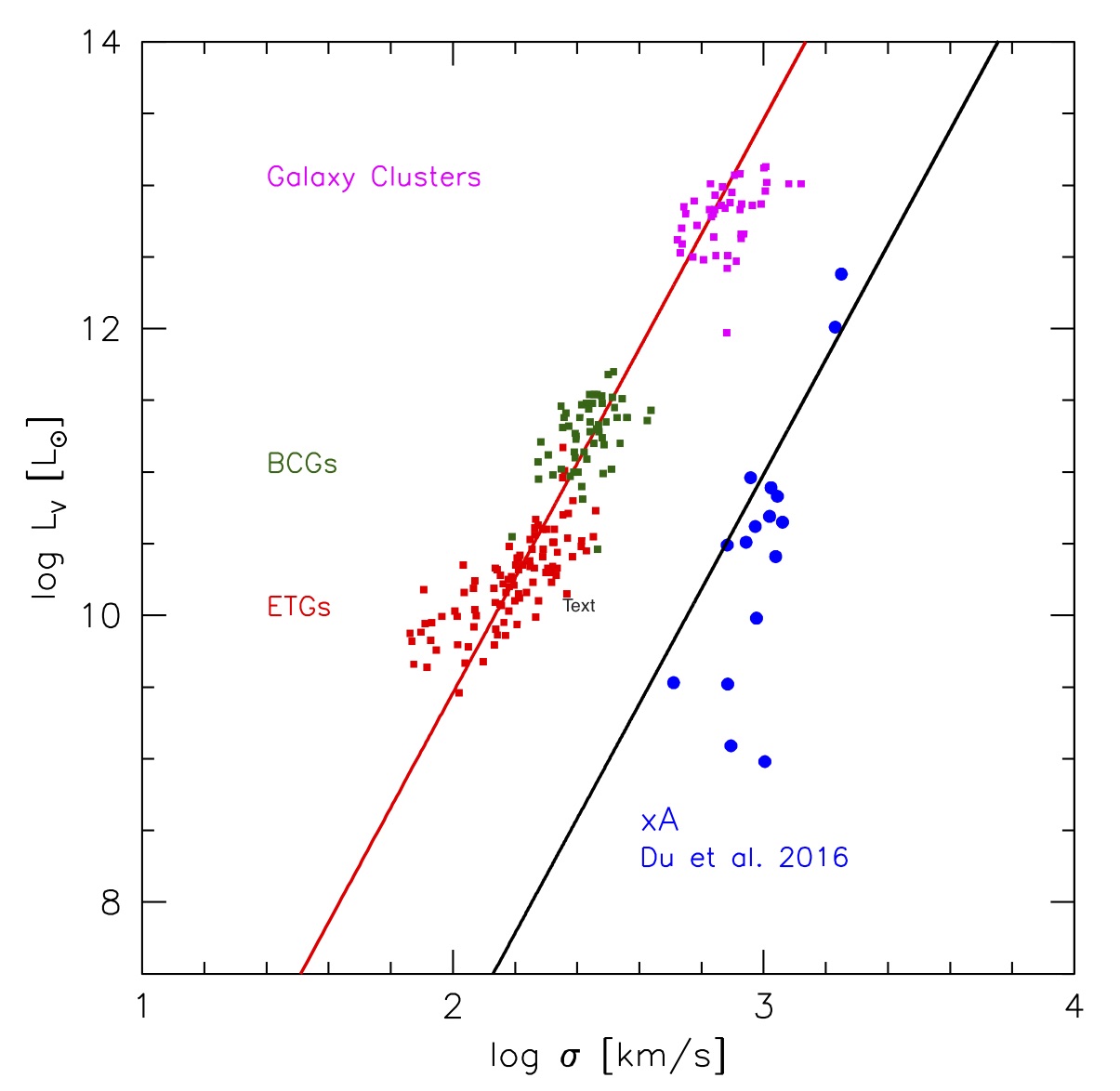}}
\caption{Relation between virial broadening and luminosity, for several classes of virialized stellar systems and quasars. {\bf Data points refers to early-type galaxies (ETGs, red squares), brightest cluster galaxies (BCGs, dark-green squares),  clusters of galaxies (magenta squares) from the wide-field nearby galaxy-cluster survey and are as used in \citet{donofrioetal19}, and xA quasars from the sample of \citet{duetal16a}.  }The filled lines trace two relations with $L \propto \sigma^{4}$.   }
\label{fig:lsigma}
\end{figure}

The $L\propto $FWHM$^{4}$\
is a law analogous to the Tully-Fisher and the early formulation of the Faber Jackson laws for early- type galaxies \citep{faberjackson76,tullyfisher77}. Galaxies and even clusters of galaxies are virialized systems that globally follows a law $\propto \sigma^{4}$ (Fig. \ref{fig:lsigma}). The main difference with quasars is that the velocity dispersion of stellar system is by far not as strongly dependent on the viewing angle as the FWHM of quasar LILs.

\section{Interpretation of the virial luminosity estimates for quasars}
\label{cs}

\subsection{Virial luminosity and redshift-based luminosity}

The sketch of Fig. \ref{fig:vw} suggests that the viewing angle of the plane of the accretion disk should substantially affect the projection of the virial velocity field along the line-of-sight. In other words, the FWHM is strongly dependent on the viewing angle $\theta$ defined as the angle between the line of sight and the axis of the accretion disk plane.

\begin{figure}[ht!]
\includegraphics[width=0.95\textwidth,clip=]{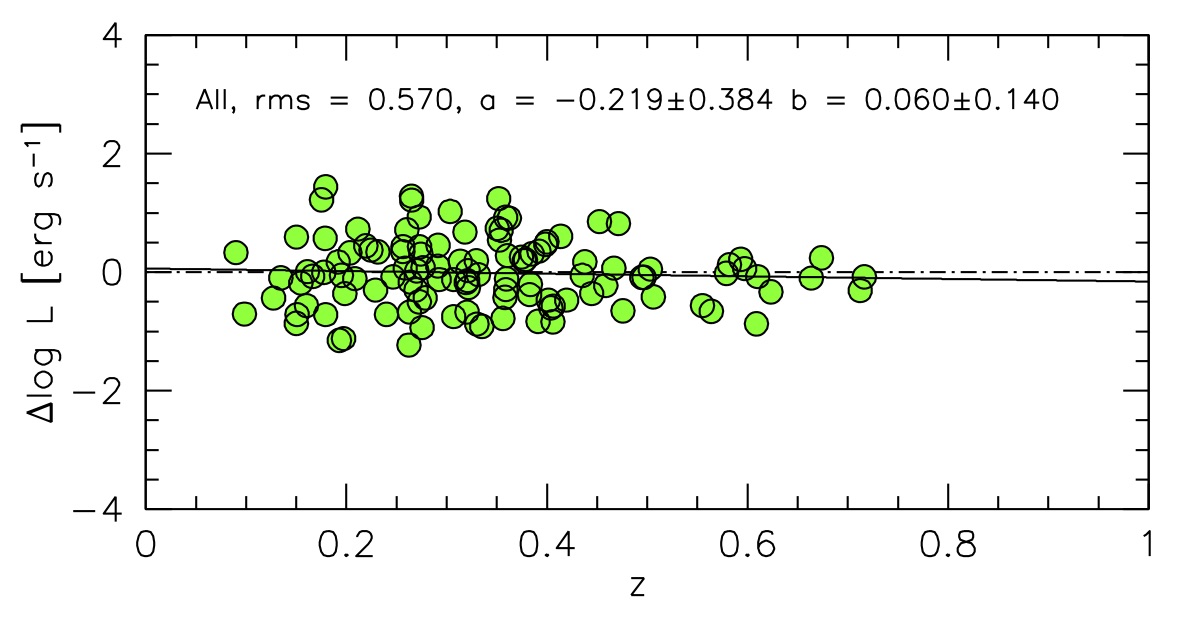}
\includegraphics[width=0.95\textwidth,clip=]{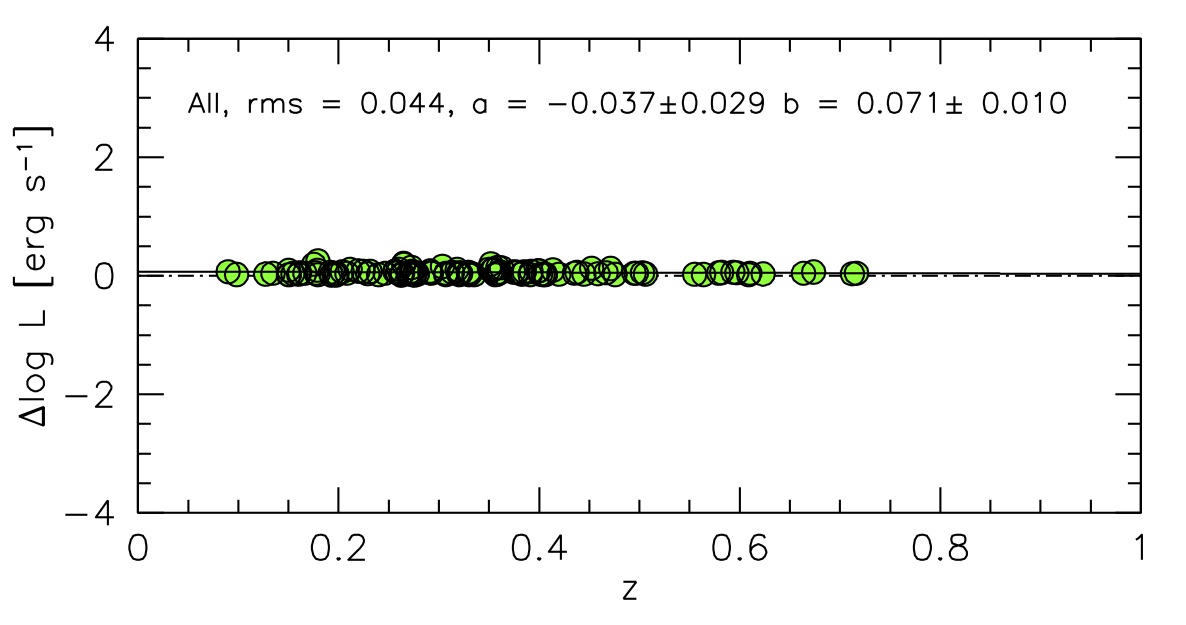}
\caption{Quasar luminosity estimates: residuals {\bf as a function of redshift $z$\ } of virial luminosity {\bf estimated from the H$\beta$ FWHM minus luminosity $L(z, H_{0}, \Omega_\mathrm{M}, \Omega_{\Lambda})$} from redshift before orientation correction (above), and after ({\bf below}). {\bf Data are from the low-$z$  quasar sample of \citet{negreteetal18}. }\label{fig:hd}}
\label{fsinus}
\end{figure}

The virial luminosity is applicable to xA quasars  over a wide range of luminosity and redshift. If we compare the virial luminosity to the conventional estimate of luminosity from redshift, $H_{0}$ and the $\Omega$s we find that there is an overall  consistency with redshift-based concordance luminosity.  At the same time we measure a significant scatter $\sigma \sim$~ 0.5 dex (Fig. \ref{fig:hd}; 0.3 dex can be reached by improving the statistics of larger samples, as done in \citealt{marzianietal17}, but a substantial statistical scatter is expected to remain even in large samples of excellent data).

An analysis aimed at the cosmological exploitation of the virial luminosity is ongoing. Work to improve the accuracy of black hole mass and Eddington ratio using $\theta$ is  also in progress.  At this point, however, it is perhaps  more interesting to try to understand the origin of the scatter between virial luminosity and conventional luminosity $L(z, H_{0}, \Omega_\mathrm{M}, \Omega_{\Lambda})$, where the cosmological parameter have been assumed to have values consistent with the concordance ones. 

The difference between the virial $L$ and $L(z, H_{0}, \Omega_\mathrm{M}, \Omega_{\Lambda})$ as a function of redshift for the sample of \citet{negreteetal18} is shown in the left panel of Fig. \ref{fig:hd}. The scatter is   due in part to measurement errors. The FWHM enters to the fourth power in the $L$\ expression; a 10\%\ error, achievable with high S/N data would imply a $\approx$0.17 dex error on luminosity (see \citealt{marzianisulentic14} for a preliminary error budget). More than measurement errors, it is likely that orientation accounts for most of the scatter, especially if lines are emitted in a flattened system. There is growing evidence that this is the case \citep{mejia-restrepoetal18,afanasievetal19,marzianietal19}   although the expression of the form factor connecting virial velocity and projected broadening is still debated. Assuming a structure factor $f$\ relating virial broadening $\delta v_\mathrm{K}$\ and line FWHM ($\delta v_\mathrm{K}^{2}= f$ FWHM$^{2}$) in the form $f = \mathrm{1}/{4(\kappa^{2} +\sin^{2}\theta )}$, where $\kappa$ can be interpreted as the ratio between an isotropic velocity component and the virial velocity, we found that all objects in the sample of \citet{negreteetal18}  can be accounted for by the effect of the viewing angle within $0 \la \theta \la 50$ {\bf degrees} (in the right panel of Fig. \ref{fig:hd}  residuals are zeroed   if  $f^{1/2}$FWHM is used as a VBE).  Orientation might be really the main source of scatter between virial and conventional luminosity estimates. More details are given by \citet{negreteetal18}, and a more conclusive analysis will be hopefully presented in forthcoming studies.   

 \section{ Conclusion}
 The MS offer contextualization of quasar observational and physical properties. Extreme Population A (xA) quasars at the high $R_\mathrm{FeII}$ end of the MS appear to radiate at extreme $L/L_\mathrm{Edd}$.  xA quasars show a relatively high prevalence (10\%) and are easily recognizable. Low ionization lines are apparently emitted in a highly-flattened, virialized BLR, and the consistency between virial and redshift-based luminosity estimates supports this basic interpretation also for xA quasars.   Several methods to derive redshift-independent $L$ values  based on intrinsic properties of quasars have been proposed in the last few years \citep{wangetal13,lafrancaetal14,risalitilusso15};   xA quasars might be   suitable as  ``Eddington standard candles''  especially if orientation effects can be accounted for.

%\begin{table}[t]
%\small
%\begin{center}
%\caption{A fraction $p_{\uparrow}M_{\rm K}$ of the mass of comets of the
%original comet formation ejected by a given planet.}
%\label{t1}
%\begin{tabular}{llll}
%\hline\hline
%planet  &{\it u}&    {\it h}      & $p_{\uparrow}M_{\rm K}$\\
%\hline
%Jupiter & 0.08  & 0.03 $\pm$ 0.01 & 2.75$nM_{\rm KP}$ \\
%Saturn  & 0.16  & 0.16 $\pm$ 0.04 & 1.16$nM_{\rm KP}$ \\
%Uranus  & 0.24  & 1.3  $\pm$ 0.5  & 0.43$nM_{\rm KP}$ \\
%Neptune & 0.52  & 2.6  $\pm$ 0.7  & 0.72$nM_{\rm KP}$ \\
%\hline\hline
%\end{tabular}
%\end{center}
%\end{table}
%%%%%%%%%%%%%%%%%%%%%%%%%%%%%%%%%%%%%%%%%%%%%%%%%%%%
%\label{f1}
%\end{figure}

\acknowledgements
PM and MDO acknowledge funding from  the INAF PRIN-SKA 2017 program 1.05.01.88.04. PM is also gtrateful for support via a STSM of the COST Action CA16104, Gravitational waves, black holes and fundamental physics that allowed her participation to the SCSLSA12. AdO acknowledges financial support from the Spanish Ministry of Economy and Competitiveness through grant AYA2016-76682-C3-1-P and from the State Agency for Research of the Spanish MCIU through the Center of Excellence Severo Ochoa award for the Instituto de Astrof\'{\i}sica de Andaluc\'{\i}a (SEV-2017-0709).  E.B. and N.B. acknowledge support from the  Ministry of Education, Science and Technological Development of the Republic of Serbia through the projects Astrophysical Spectroscopy of Extragalactic Objects (176001) and Gravitation and structure of the Universe on large scales  (176003).

%\bibliography{biblioletter2a}

\end{document}